\documentclass[pra,twocolumn,showpacs,superscriptaddress,longbibliography,floatfix]{revtex4-1}

\usepackage{lmodern} %Latin Modern = Extended version of Computer modern
\usepackage[T1]{fontenc}
\usepackage[utf8]{inputenc}
\usepackage{microtype}

\usepackage{amssymb,amsmath,braket,cancel}
\usepackage{siunitx}
\usepackage{commath} % includes \dif \abs etc

\usepackage{url}

\usepackage[
    ]{todonotes}

\usepackage{graphicx}

\begin{document}

\title{Enhanced spectral sensitivity of a chip-scale photonic-crystal slow-light interferometer}

\author{Omar S. Maga\~{n}a-Loaiza}
\affiliation{The Institute of Optics, University of Rochester, Rochester, New York 14627, USA}

\author{Boshen Gao}
\affiliation{The Institute of Optics, University of Rochester, Rochester, New York 14627, USA}

\author{Sebastian A. Schulz}
\affiliation{Department of Physics and Max Planck Centre for Extreme and Quantum Photonics, University of Ottawa, 25 Templeton, Ottawa, ON, K1N 6N5, Canada}

\author{Kashif Awan}
\affiliation{School of Electrical Engineering and Computer Science, University of Ottawa, 25 Templeton, Ottawa, ON, K1N 6N5, Canada}

\author{Jeremy Upham}
\affiliation{Department of Physics and Max Planck Centre for Extreme and Quantum Photonics, University of Ottawa, 25 Templeton, Ottawa, ON, K1N 6N5, Canada}

\author{Ksenia Dolgaleva}
\affiliation{School of Electrical Engineering and Computer Science, University of Ottawa, 25 Templeton, Ottawa, ON, K1N 6N5, Canada}

\author{Robert~W.~Boyd}
\affiliation{The Institute of Optics, University of Rochester, Rochester, New York 14627, USA}
\affiliation{Department of Physics and Max Planck Centre for Extreme and Quantum Photonics, University of Ottawa, 25 Templeton, Ottawa, ON, K1N 6N5, Canada}

\date{\today}

\begin{abstract}

We experimentally demonstrate that the spectral sensitivity of a Mach-Zehnder (MZ) interferometer can be enhanced through structural slow light. We observe a 20 times enhancement by placing a dispersion-engineered-slow-light photonic-crystal waveguide in one arm of a fibre-based MZ interferometer. The spectral sensitivity of the interferometer increases roughly linearly with the group index, and we have quantified the resolution in terms of the spectral density of interference fringes. These results show promise for the use of slow-light methods for developing novel tools for optical metrology and, specifically, for compact high-resolution spectrometers.

\end{abstract}

%\pacs{03.65.Ta, 42.50.Tx, 42.50.Ex, 42.25.Hz }
\maketitle

Slow light has fascinated the physics community for over two decades \cite{Boyd:2009th, Boyd:2009hf}. The slowdown of light has been observed in a diverse range of media including atomic vapors, optical fibers, and photonic crystals (PhCs) \cite{Boyd:2011}. The ability to manipulate the speed of light has led to a wide variety of technological applications, for example, optical buffers, optical memories, laser radars, and enhanced spectrometers \cite{Boyd:2009th, Boyd:2009hf, Baba:2008ks, Krauss:2007je,Krauss:2008tf, schweinsberg:2011jw, Shi:2008he}. 

Optical interferometry has benefited greatly from the use of highly dispersive materials \cite{Soljacic:2002bd, Shi:2007gn, Shi:2007un, Wang:2011fz, Marandi:2011md}. For instance, it has been predicted that a Mach-Zehnder (MZ) interferometer with a slow-light medium in one of the arms will greatly enhance spectral sensitivity with respect to its conventional (non-slow-light) version \cite{Soljacic:2002bd, Shi:2007gn, Shi:2007un}. This result holds promise for compact and efficient spectrometers \cite{Boyd:2009th, Shi:2008he}. Nevertheless, up to this date, the improvement of the spectral sensitivity of interferometers has been demonstrated only in bulk optical setups, such as shearing and free-space interferometers, with highly dispersive crystals and atomic vapors \cite{Shi:2007gn, Shi:2007un, Wang:2011fz}. 

\begin{figure}[ht]
\begin{center}
 \includegraphics[width=0.47\textwidth]{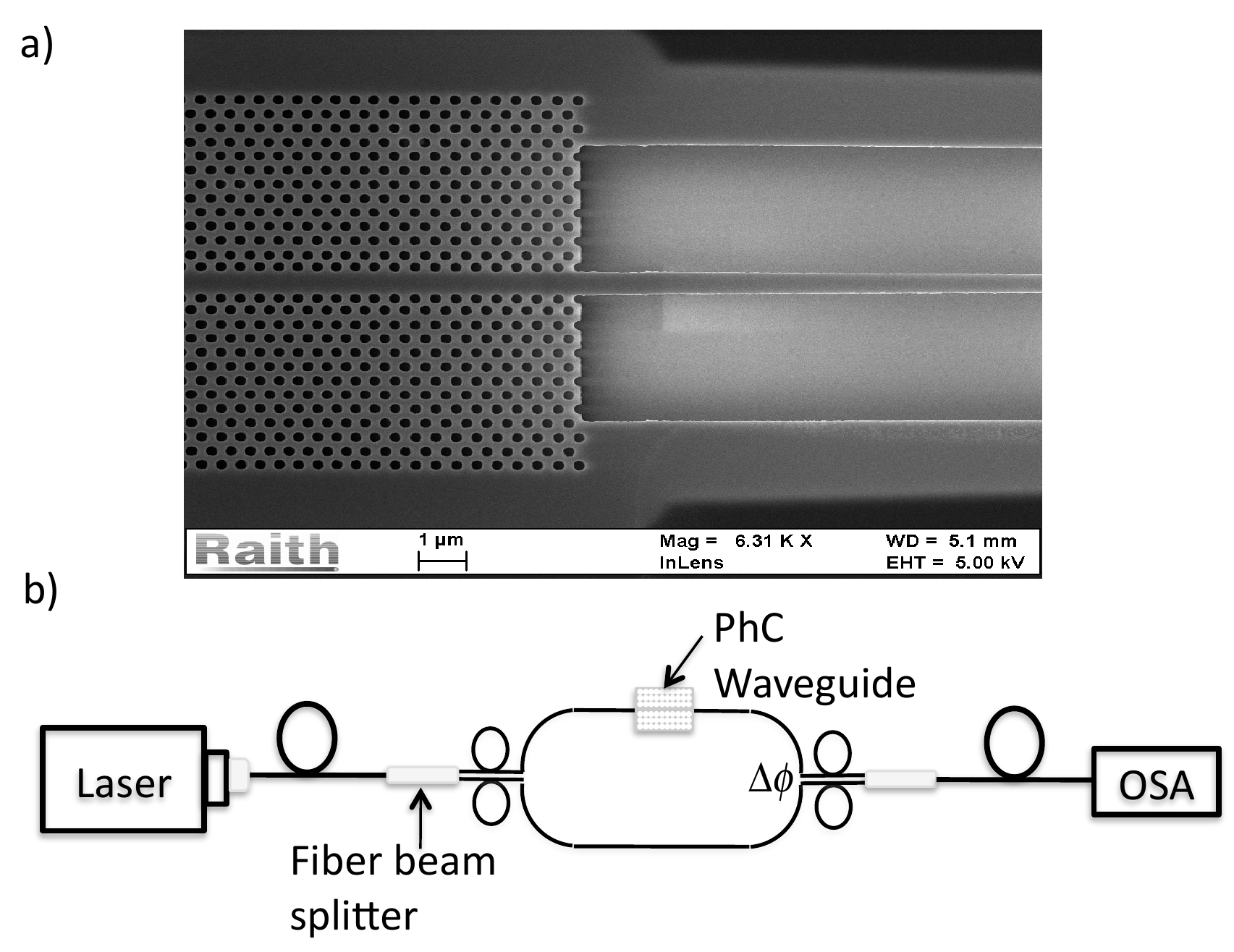}
 \caption{Slow-light waveguide and measurement setup. a) Scanning electron microscope (SEM) photograph of the fabricated PhC waveguide and its access channel. b) Fiber-based Mach-Zehnder interferometer containing a PhC slow-light  waveguide in one of the arms. The difference in optical pathlength produced by the imbalance between the two arms produced by the slow-light region is indicated by $\Delta\phi$. 
 }
 \end {center}
\label{fig:fig1}
\end{figure}
While not achieving the same extreme slow-down values as material slow light, slow light in nanophotonic devices, particularly in photonic crystal (PhC) waveguides, has many key advantages. Nanophotonic slow light structures are extremely compact, with low footprint and high mechanical stability compared to material slow light systems \cite{Boyd:2011}. Furthermore, the availability of high-precision fabrication techniques and the use of common materials, e.g. silicon, makes nanophotonic slow light particularly interesting for applications \cite{Boyd:2011,Baba:2008ks, Krauss:2007je, Krauss:2008tf}. Here we demonstrate that structural slow light can dramatically improve the spectral sensitivity of an interferometer. Thus, we combine the enhancement to the spectral sensitivity with the advantages of nanophotonics. Specifically, we place a silicon PhC slow-light waveguide within a compact fiber-based MZ interferometer; see Fig. 1. Our results show an enhancement of a factor of 20 in the spectral sensitivity of the interferometer, without increasing its physical size. This sensitivity is on par with that of previous implementations of material slow-light interferometers \cite{Shi:2007gn}, using a much more compact slow light structure (the total PhC footprint is on the order of 0.015mm$^2$)  \cite{Shi:2007gn, Shi:2007un, Wang:2011fz}.
\begin{figure*}[ht]
\begin{center}
 \includegraphics[width=1\textwidth]{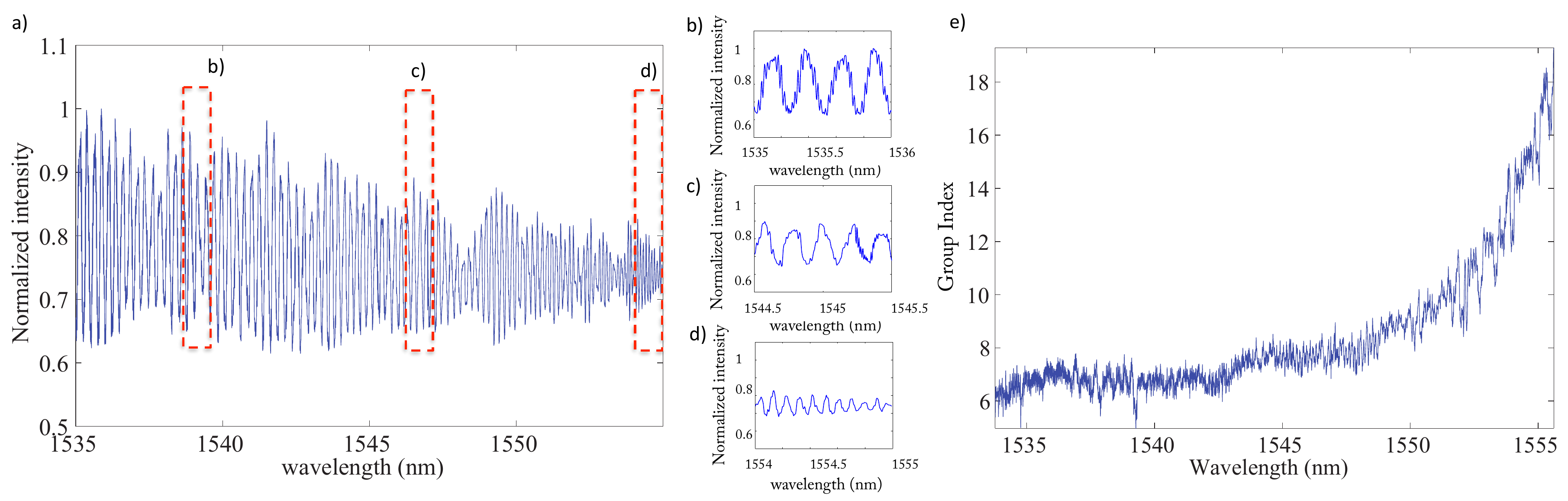}
 \caption{ (a) Intensity of one of the outputs of the MZ interferometer plotted as a function of wavelength.  Note that the fringe density increases with increasing wavelength. Insets  (b) through (d) show three different 1 nm spans with significantly different fringe spacings. (e) The wavelength-dependence of the group index. }
 \end {center}
\label{fig:fig2}
\end{figure*}
In order to demonstrate the advantage of adding a slow-light medium into a spectral interferometer, we discuss the phase difference $\Delta\phi$ between two light beams passing through the two arms of a MZ interferometer. We assume that the phase difference is solely caused by propagation through a length $L$ of slow-light material. The phase difference is defined as
\begin{equation}
\Delta\phi=\frac{\omega}{c} n(\omega) L,
\end{equation}
\\
\noindent
where $\omega$ is the frequency of the propagating light, $c$ the speed of light in vacuum, and $n(\omega)$ the refractive index of the medium. In practice, $\Delta\phi$ is determined by performing intensity measurements of the interference pattern produced at one of the output ports of the interferometer:
\begin{equation}
I=\frac{I_0}{2}(1+\cos{\Delta\phi}),
\end{equation}
\\
\noindent
where $I_0$ is the input intensity. The spectral resolution of the interferometer is determined by the minimum resolvable change in intensity due to a shift in frequency \cite{Dowling:2008hs}. This change is quantified by the first derivative of the intensity with respect to the frequency:
\begin{equation}
\frac{dI}{d\omega}=\frac{I_0}{2}\frac{d\Delta\phi}{d\omega}\sin{\Delta\phi}=\frac{I_0}{2}\frac{L}{c}\left(n+\omega\frac{dn}{d\omega}\right)\sin{\Delta\phi}.
\end{equation}
\\
\noindent
We note that $n+\omega\ {dn}/ {d\omega}$ is the definition of the group index, $n_g$. Thus Eq.\ (3) can be rewritten as
\begin{equation}
\frac{dI}{d\omega}=\frac{I_0}{2}\frac{Ln_g}{c}\sin{\Delta\phi}.
\end{equation}
\\
\noindent
We see that a larger group index leads to denser fringes,
which in turn leads to the enhancement of the spectral sensitivity
of the interferometer \cite{Dowling:2008hs}. This enhancement 
depends on the product $n_gL$. Thus using a slow light medium with a high $n_g$ reduces the physical size of the interferometer needed to achieve a given spectral sensitivity.
\begin{figure}[ht]
\begin{center}
 \includegraphics[width=0.45\textwidth]{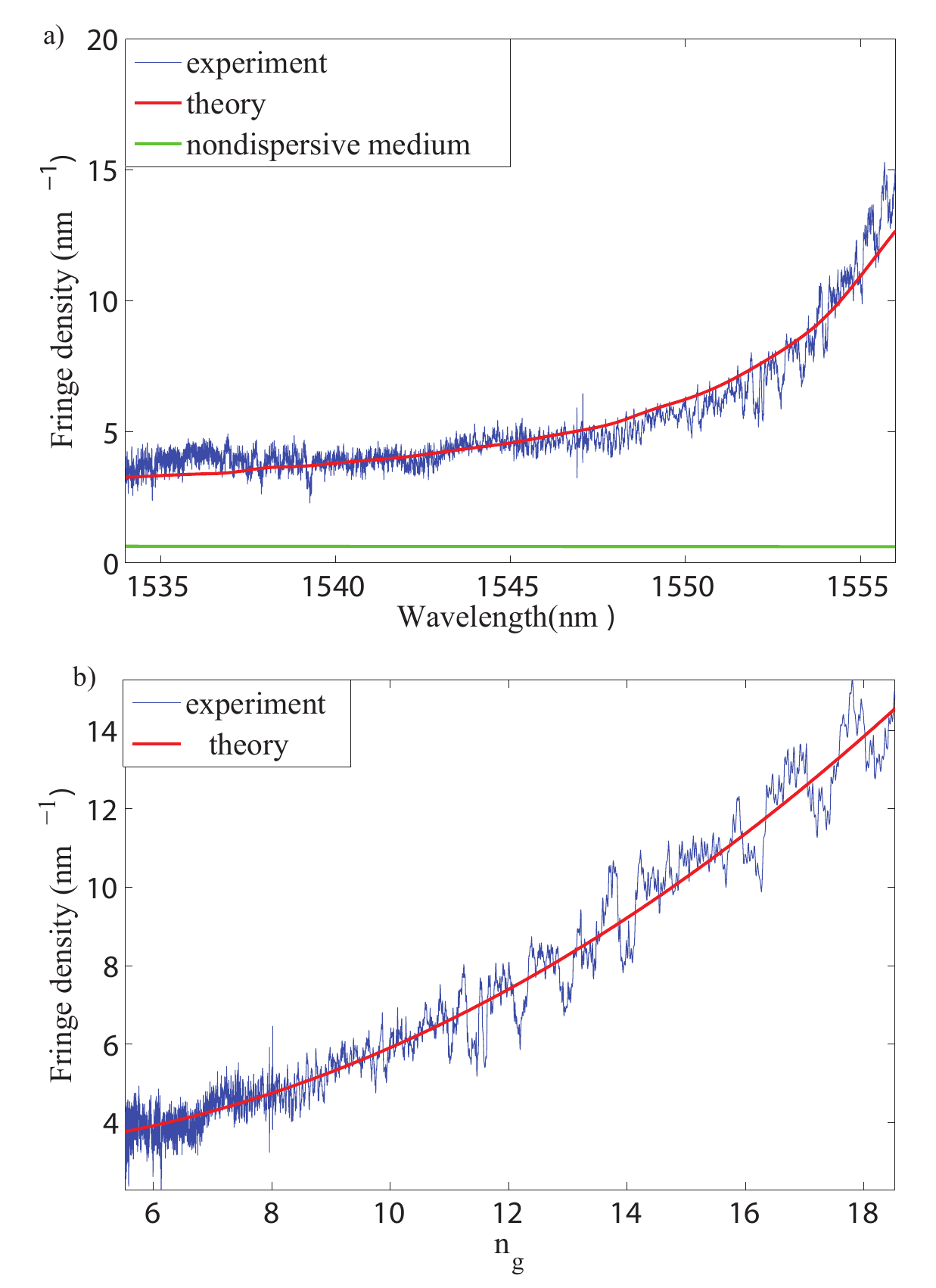}
 \caption {(a) Experimental measurements and theoretical predictions of the density of interference fringes plotted as a function of wavelength for the slow-light MZ interferometer of Fig.\ 1 (a). (b) The dependance of fringe density on group index $n_g$ for the same waveguide. We see that the fringe density, and consequently the spectral sensitivity, increases with the group index of the slow-light waveguide.  
 }
 \end {center}
\label{fig:fig3}
\end{figure}
The group index, $n_g$, can be significantly higher than the refractive index, $n$, of a material and therefore slow light materials can significantly enhance the spectral sensitivity of interferometers \cite{Boyd:2009th, Boyd:2009hf, Shi:2008he}. In our experiment we measure the interference pattern formed at the output of the MZ interferometer as a function of wavelength $\lambda$. This measurement allows us to quantify the enhancement in resolution and sensitivity through an experimental parameter, the density of fringes per unit wavelength interval, $N_{\lambda}$. An approximate expression for the local density of fringes over a small frequency range $\Delta\omega$ can be derived by describing the refractive index of the dispersive material as $n \approx n_0 + (n_g(\lambda)-n_0)\Delta\omega$, where it is assumed that $n_g$ remains approximately constant over the narrow frequency range being considered. This approximation and the relation between wavenumber and frequency allows us to define the local fringe density as:
\begin{equation}
N_{\lambda}=\frac{\sqrt{n_0^2L^2+8\{n_g(\lambda)-n_0\}\pi c L}+n_0L}{2\lambda^2}.
\end{equation}
\\
\noindent
For a non-dispersive medium ($n_g=n_0$), this equation recovers the result for a traditional free-space interferometer: $N_{\lambda}={n_0L}/{\lambda^2}$. For a slow-light medium, where $n_g > n_0$, the fringe density is  increased. Thus, from Eq. 5, we expect a reduction in the fringe density with increasing wavelength and an increase in the fringe density with increasing group index. We recall that the group index itself is strongly wavelength dependent and increases monotonically with wavelength in the spectral region considered here.

We engineered the dispersive properties of our W1-based PhC waveguides through a shift of the first and second row of holes perpendicular to the waveguide direction \cite{Li:2008} and include coupling interfaces to maximize the transmission \cite{Schulz:2010}. The waveguide has a lattice constant of 416 nm, a hole radius of 124 nm, the silicon slab is 210 nm thick, and the first and second rows are shifted by 46 nm outwards and 16 nm inwards, respectively. The waveguide was fabricated according to \cite{Awan:2015fz}. A scanning electron microscope (SEM) image of the device is shown in Fig.\ 1 a). 

The slow light characteristics of the PhC waveguide are of fundamental importance to this study and were determined through Fourier spectral interferometry \cite{GomezIglesias:2007}. Due to the nature of our slow-light material and its mechanical stability, we chose a fiber-based MZ interferometer with the slow-light waveguide in one of the arms. This setup is used as a proof-of-principle experiment to demonstrate the enhancement of the sensitivity due to structural slow light, see Fig.\ 1 b). Light is coupled into and out of the waveguide using grating couplers that were etched at the same time as the PhC waveguide, see Fig.\ 1 a). The optical output of the slow-light MZ interferometer is measured as a function of wavelength through the use of an optical spectrum analyzer (OSA).  We use a broadband, amplified spontaneous emission (ASE) source spanning the wavelength range from 1525 nm to 1575 nm. The obtained interference pattern is shown in Fig.\ 2 a)-d). The modulation of the fringe height is due to a resonant effect caused by reflections from the grating couplers of the PhC sample and is not intrinsic to a slow light interferometer. 
The increase in fringe density with wavelength is a consequence of the slow light effect in the PhC waveguide. The experimental measurement, Fig.\ 2 e), shows the increase in group index as a function of wavelength. As predicted by Eqs.\ 4 and 5, when the dispersive properties of the material become significant, they lead to an enhancement of the resolution and sensitivity of the device.  

In Fig\ 3 a) we show the experimental characterization of the enhancement in sensitivity of our interferometer. We plot the fringe density of the slow-light MZ interferometer as a function of wavelength. The increased density at longer wavelength leads to an increase in the spectral resolution of the device. The red curve in Fig.\ 3 a) shows the fringe density predicted by our numerical simulation using a plane wave expansion approach (MPB \cite{Johnson2001:mpb}). Further, the expected fringe density of a traditional ($n_g \approx n$) MZ interferometer is plotted in green. Clearly, there is a dramatic improvement in the fringe density, and therefore in the sensitivity of the interferometer, when a slow-light structure is incorporated within it.

For this particular PhC waveguide, our proof-of-principle experiments shows a 20 times increase in the sensitivity. This number is taken from Fig.\ 3 a) by calculating the ratio between the fringe density for the interferometer with (curve in red) and without (curve in green) the dispersive medium at a wavelength of 1556 nm. Further evidence of the role of the group index in determining the sensitivity of the interferometer is shown in Fig.\ 3 b). In this case, the increased resolution of the device, quantified by its fringe density, exhibits a direct relationship with the group index, as described by Eqs.\ 3 - 5. This group index dependance allows one to keep the physical size of interferometers small, creating the potential for compact spectrometers. The spectral sensitivity could be optimized for different applications by engineering the PhC waveguide properties. For instance, further increasing the resolution or creating spectral ranges with a constant enhancement.

In summary, we have demonstrated an enhancement of the spectral resolution, accompanied by enhanced sensitivity, of a MZ interferometer by incorporating a structural slow light component, in our case a PhC waveguide. This enhancement, which reached a factor of 20 in our experiment, can be further improved on through careful waveguide design. We demonstrated the increased sensitivity through a significant increase in the density of interference fringes. Enhanced-sensitvity interferometers have been an important subject of interest in the photonics community for the last ten years \cite{Boyd:2009th, Boyd:2009hf, Baba:2008ks, Krauss:2007je,Krauss:2008tf, Shi:2008he, Soljacic:2002bd, Shi:2007gn, Shi:2007un, Wang:2011fz, Marandi:2011md}. We therefore believe that our experimental results are an important starting point for further studies, including on-chip implementations of our device. These results provide a path towards a new generation of compact sensitive devices and sensors utilizing light.  

\section{Acknowledgments}

The authors would like to thank David D. Smith,
and Jerry Kuper for many fruitful discussions. This work was supported by the US Defense
Threat Reduction Agency Joint Science and Technology Office for Chemical and Biological
Defense (Grant No. HDTRA1-10-1-0025), by the National Aeronautics and Space Administration
(under contract NNX15CM47P), and by the Canada Excellence Research Chairs Program.

\bibliography{Refs}
\end{document}